\documentstyle[twoside,fleqn,npb,epsfig]{article}

\newcommand{\ket}{\rangle}
\newcommand{\eg}{{\it e.g.\ }}
\newcommand{\ie}{{\it i.e.\ }}

\title{A model for parton distributions in hadrons\thanks{Contribution to
the DIS99 workshop proceedings}}

\author{A. Edin%
        \address{DESY, Notkestrasse 85, DE-22603 Hamburg, Germany}%
        \thanks{Anders.Edin@desy.de}
        and
        G. Ingelman%
        \address{DESY, Notkestrasse 85, DE-22603 Hamburg, Germany and \\
	Uppsala University, Box 535, SE-751 21 Uppsala, Sweden}%
        \thanks{ingelman@desy.de}
}

\begin{document}

\begin{abstract}
The non-perturbative parton distributions in hadrons are derived from simple 
physical arguments resulting in an analytical expression for the valence 
parton distributions. The sea partons arise mainly from pions in hadronic 
fluctuations. The model gives new insights and a good description of 
structure function data.
\end{abstract}

\maketitle
\vspace*{-75mm}
\noindent
TSL/ISV-99-0213    
\vspace*{65mm}

Hard processes involving hadrons are calculated by folding perturbative QCD
matrix elements with parton distributions describing the probability  of
finding a quark or a gluon in the hadron.

Perturbative QCD evolution describes the dependence of the parton distributions
on the hard scale $Q$ of the interaction. However, their dependence on the
momentum fraction $x$ at the lower limit for applying perturbative QCD,
$Q_0\approx 0.5-2$ GeV, are fitted to data using parameterisations, \eg of the
form
\cite{CTEQ5,GRV98,MRS}
\begin{equation}
f_i(x,Q_0)=N_i x^{a_i} (1-x)^{b_i} (1+c_i \sqrt{x} + d_i x)
\end{equation}
The parameters in these functions have no direct physical meaning, making it
difficult to interpret the results. To gain understanding of non-perturbative
QCD we have developed  a physical model \cite{EI_partons98} for the parton
distributions at $Q_0$.  This is here briefly described together with our
latest developments. 

\begin{figure}[tb]
\center{\epsfig{width=0.6 \columnwidth,file=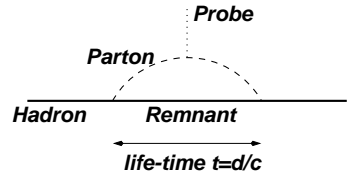}}
\vspace*{-8mm}
\caption{Fluctuation of a hadron into a parton and a remnant.}
\label{fig:fluctuation}
\end{figure}

The basic physical picture is that a probe with large resolution,  compared to 
the hadron size, will see {\it free quarks and gluons} in quantum fluctuations 
of the hadron. The measuring time is short compared to the life-time of the
fluctuation since the latter is determined by the confinement of quarks and
gluons inside the hadron, as illustrated in Fig.~\ref{fig:fluctuation}. This
makes it possible to describe the formation of the fluctuations independently
of the measuring process.

\begin{figure}[tb]
\center{\epsfig{width=0.7 \columnwidth,file=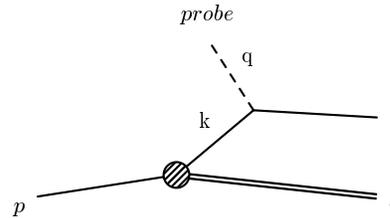}}
\vspace*{-5mm}
\caption{Kinematics for scattering on a parton in a hadron.}
\label{fig:kin}
\end{figure}

Our approach only intends to provide the four-momentum $k$ of a single probed 
parton. All other information in the hadron wave function is neglected,
treating the other partons collectively as a remnant with four-momentum $r$,
see Fig.~\ref{fig:kin}.

It is convenient to describe the process in the hadron rest frame where  there
is no preferred direction and hence spherical symmetry.  The probability
distribution for finding one parton is taken as a Gaussian,  which expressed in
momentum space for a parton with four-momentum $k$ and mass $m_i$ is 
\begin{equation}
f_i(k)dk=N(\sigma_i,m_i)
          e^{-\frac{(k_0-m_i)^{2}+k_1^2+k_2^2+k_3^2}{2\sigma_i^2}}dk,
\label{eq:fk}
\end{equation}
where $\sigma=\frac{1}{d_h}\approx m_{\pi}$ is the inverse of the confinement
length scale or hadron diameter.

The partonic structure is described using the light-cone momentum fraction 
$x=\frac{k_+}{p_+}$ which the parton has in the initial hadron. Since $x$ is
invariant under boosts in the $z$ direction, the same will be true for the
calculated parton distributions. 

There are a number of constraints that must be fullfilled by the parton
distributions. The normalisation for valence quarks is given by the sum
rules
\begin{equation}
\int_0^1 f_i(x)dx = n_i, 
\end{equation}
and for the gluons by the momentum sum rule
\begin{equation}
\sum_i\int_0^1 xf_{i}(x)dx = 1.
\end{equation}

There are also the kinematical constraints
\begin{equation}
m_{i}^{2}\le j^{2}<W^{2} \;\;\; {\rm and} \;\;\; r^{2}>\sum _{i}m_{i}^{2} 
\label{eq:constraints}
\end{equation}
given by the final partons being on-shell or time-like and the remnant having
to include the remaining partons.  These constraints also leads to $0<x<1$.

The parton model requires that $W$ is well above the resonance region and 
that the resolution of the probe is much larger than the size of the hadron, 
\ie 
\begin{equation}
W \gg m_{p} \;\;\; {\rm and} \;\;\; Q_0 \gg \sigma _{i}
\end{equation}

The scale of the probe must also be large enough, $Q_0\gg \Lambda _{QCD}$,  for
perturbative QCD to describe the evolution of the parton distributions from the
starting scale $Q_0$.

In~\cite{EI_partons98} we integrated Eq.~\ref{eq:fk} numerically 
to find the parton distributions since the kinematical 
constraints, Eqs.~\ref{eq:constraints}, are quite complicated in general. 
The problem is much simpler if the transverse momenta and the masses of the
partons are neglected. It is then possible to derive \cite{E_partons99}
an analytical expression for the parton distributions,  
\begin{equation}
f_i(x)=N'(\tilde{\sigma}_i)\exp \left( -\frac{x^2}{4\tilde{\sigma}_i^2}\right) 
{\rm erf}\left( \frac{1-x}{2\tilde{\sigma}_i}\right),
\label{eq:fx}
\end{equation}
where
\begin{equation}
\tilde\sigma = \frac{1}{d_h m_h} \approx \frac{m_{\pi}}{m_h}.
\end{equation}

\begin{figure}[tb] 
\center{  
   \epsfig{width=0.8 \columnwidth,file=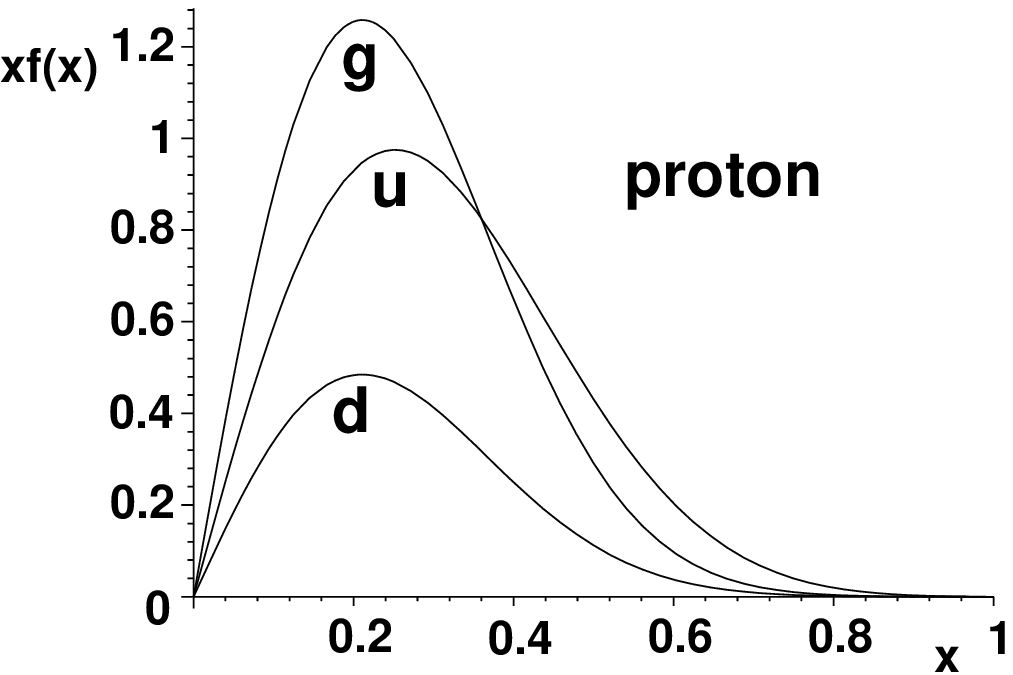} 
\vspace*{5mm}\\
   \epsfig{width=0.8 \columnwidth,file=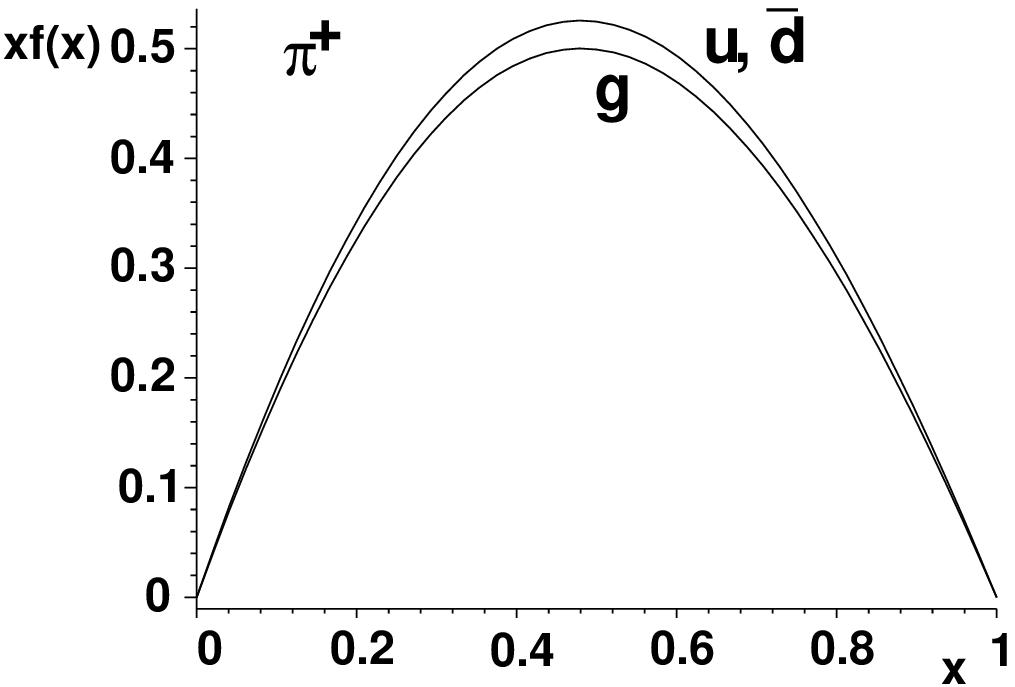}
       } 
\vspace*{-8mm}
\caption{Valence distributions at $Q_0\approx 0.85$ GeV in the proton and $\pi^+$.} 
\label{fig:proton_pion} 
\end{figure} 

The valence parton distributions for hadrons are here determined simply by 
the mass and size of the hadron! The resulting valence distributions for 
the proton ($\tilde\sigma\approx 0.15$) and the pion ($\tilde\sigma\approx 1$)
are very reasonable as shown in Fig.~\ref{fig:proton_pion}. 
Note that the pion distributions are very similar to $xf(x)=2x(1-x)$ and that 
one third of the pion momentum is carried by gluons.

\begin{figure*}[t]
\center{
\epsfig{height=6cm,file=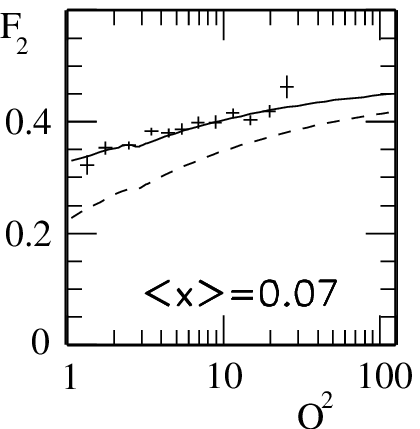}
\hspace{15mm}
\epsfig{height=6cm,file=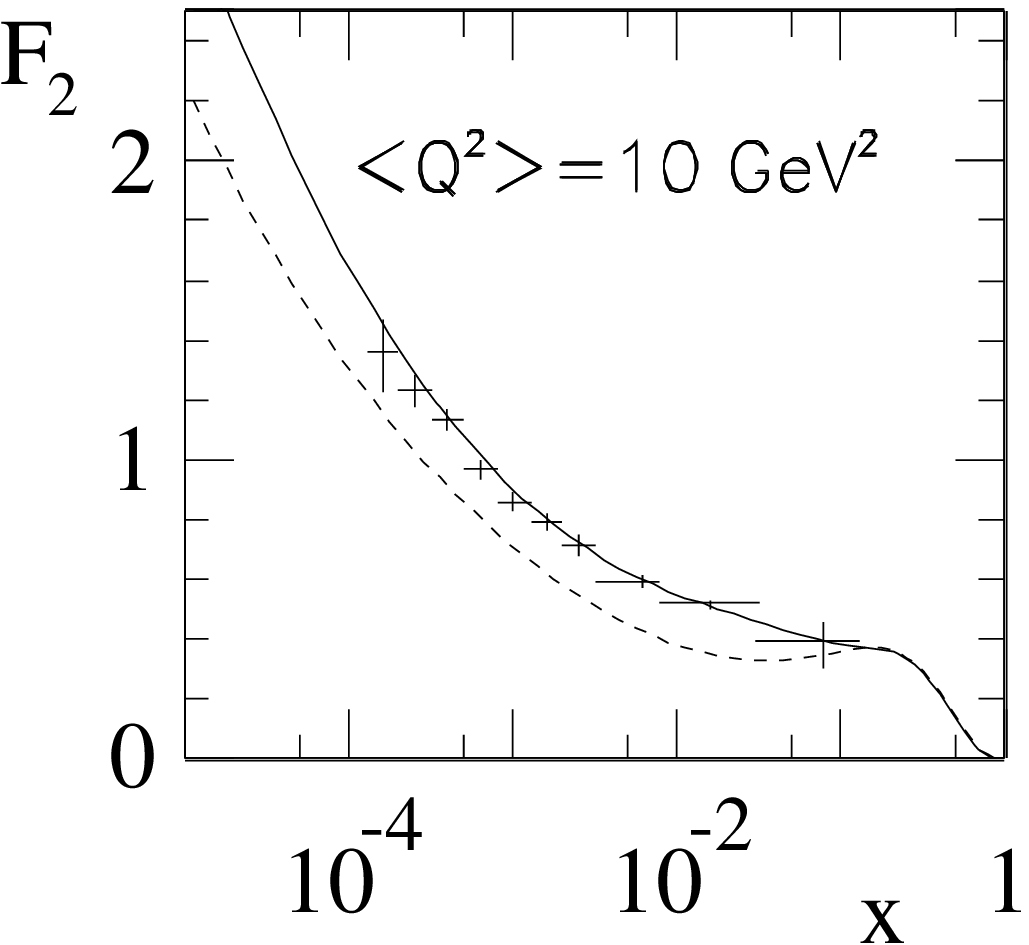}}
\vspace*{-6mm}
\caption{Data from NMC \cite{NMC} and ZEUS \cite{ZEUS} on the proton 
structure function $F_2(x,Q^2)$ compared to the model with valence partons 
only (dashed) and including sea partons (full).}
\label{fig:nmc}
\end{figure*}

The sea partons are described by hadronic fluctuations, \eg for the proton
$|p\pi^0\ket+|n\pi^+\ket+...$,  where the probe measures a valence parton in
one of the two hadrons. The momentum distribution of pions in a hadronic
fluctuation is assumed to follow from the same model as for the valence
partons, with the differences that the mass can not be neglected and the width
$\sigma_{\pi}\approx 50$ MeV is smaller, related to the longer range of pionic
strong interactions.

Using these valence and sea parton $x$ distribut\-ions at $Q_0=0.85$ GeV, 
next-to-leading order DGLAP evolution in the CTEQ program \cite{CTEQ5} was
applied to obtain the parton distributions at larger $Q$. The proton structure
function $F_2(x,Q^2)$ can then be calculated and compared with deep inelastic
scattering data, as illustrated in Fig.~\ref{fig:nmc} and detailed in
\cite{EI_partons98}. The model does remarkably well, in view of its simplicity
and few parameters (6).  Of course, conventional parton density
parameterisations give much better fits, probably mainly due to their many more
parameters ($\approx 20$). 

The main part of the proton structure is determined by the valence
distributions, but the sea gives an important contribution at small $x$, as can
also be seen from the comparison with the measured $F_2$. 

Including all pion fluctuations will give a flav\-our asymmetric sea with
$\bar{d}>\bar{u}$ as also observed experimentally \cite{Sea_asymmetry}, but the
numerical details remain to be investigated.

The model predicts the valence parton distributions for all hadrons, 
but heavy quarks gives more complicated analytical expressions 
\cite{E_partons99}. Numerical results on strange and charmed mesons 
are shown in \cite{EI_partons98}. In addition, a study based on Monte Carlo 
has been made \cite{K_Xjobb} to investigate intrinsic strange and charm quarks 
in the proton.


\begin{thebibliography}{9}
\bibitem{CTEQ5} H.L.\ Lai et al., Phys.\ Rev.\ D55 (1997) 1280 \\
                H.L.\ Lai et al., hep-ph/9903282
\bibitem{GRV98} M.\ Gl\"uck et al., Eur.\ Phys.\ J.\ C5 (1998) 461
\bibitem{MRS} A.D.\ Martin et al., Eur.\ Phys.\ J.\ C4 (1998) 463
\bibitem{EI_partons98} A.\ Edin, G.\ Ingelman, Phys.\ Lett.\ B432 (1998) 402
\bibitem{E_partons99} A.\ Edin, in preparation
\bibitem{NMC} M.\ Arneodo et al., NMC coll., Phys.\ Lett.\ B364 (1995) 107 
\bibitem{ZEUS} M.\ Derrick et al., ZEUS coll., Z.\ Phys.\ C72 (1996) 399
\bibitem{Sea_asymmetry} J.C. Peng et al., FNAL E866 coll.,  Phys.\ Rev.\ D58
(1998) 92004
\bibitem{K_Xjobb} K.\ Torokoff, Master thesis, Uppsala University, 
TSL/ISV-99-0204
\end{thebibliography}
\end{document}